 \documentclass[12pt,preprint]{aastex}

\usepackage{graphicx}
\usepackage{natbib}
\def \be {\begin{equation}}
\def \en {\end{equation}}

\def \mt {}

\def \mtt { }

\def \af{}

\def \mtg {  }

\def \af {}
\def \aff {}
\def \mtg {}
\def \mgg {}
\def \mmm {}
\def \mmmm {}
\def \mmmmn {}
\def \alf {}
\def \hh {}
\def \qq {}
\def \mn {}
\def \mnn {}

\def \pg {PG~1553+113 }
\def \pgp {PG~1553+113}

\slugcomment{Submitted to \textit{ApJ}: July 27, 2016; accepted:
Jan. 8, 2017}
\shorttitle{...}

\shortauthors{....}

\begin{document}

% \title{How Reflecting Structures Contribute to Uncorrelated Gamma-Ray
% Flares in Blazars}

% \title{Gamma-Ray flares in Blazars: Missing and Unexpected Correlations
% with Other  Bands}

\title{\large
 Blazar Jets Perturbed by Magneto-Gravitational Stresses in
Supermassive Binaries}

%The Blazar PG 1553+113 modelled as a Supermassive
% Binary:  Optical/Gamma-Ray Emissions and  X-Ray Outbursts}

\vspace{2.cm}
 \author{\large
A.~Cavaliere\altaffilmark{1,2},
 M.~Tavani\altaffilmark{1,2,3,4}, V.~Vittorini\altaffilmark{2}}

\altaffiltext{1}{Astronomia, Accademia Nazionale dei Lincei, via
della Lungara 10, I-00165 Roma, Italy}
\altaffiltext{2}{INAF/IAPS--Roma, via del Fosso del Cavaliere
 100, I-00133 Roma, Italy}
 \altaffiltext{3}{Universit\'a ``Tor Vergata'', Dipartimento di
 Fisica,
 via della Ricerca Scientifica 1, I-00133 Roma, Italy}
 \altaffiltext{4}{Gran Sasso Science Institute, viale Francesco
 Crispi 7, I-67100 L'Aquila, Italy}

%  \altaffiltext{4}{INAF/IASF--Palermo,
% Via Ugo La Malfa 153, I-90146 Palermo, Italy}}
% \altaffiltext{*}{Email: \texttt{marco.tavani@inaf.it}

\vskip .1in

%\vspace{4cm}

\begin{abstract}

We study particle acceleration and radiative processes in Blazar
jets under recurring % perturbations induced
{\mn conditions set} by gravitational % dynamics
{\mn perturbations } in supermassive binary systems.
% We model such a behavior in terms of
{\mn We consider} the action from a companion orbiting a primary
%super-massive
black hole of $\sim 10^8 \, M_{\odot}$, and
perturbing its relativistic jet. We discuss how such conditions
 induce repetitive magneto-hydrodynamic stresses along
the jet, and affect its inner electron acceleration and radiative
processes. {\mn Specifically, we {\mn study} how macroscopic
perturbations {\mn related to increased} jet "magnetization" {\mn
end up into higher} radiative outputs in the optical, X-ray and
gamma-ray bands}. We {\mn find} first an increase in magnetic
field strength as gauged in the optical band from the Synchrotron
emission of electrons accelerated in kinetic processes stimulated
by reconnecting magnetic lines. The energetic electrons then
proceed to up-scatter the Synchrotron photons to GeV energies {\mn
after} the canonical Synchrotron-Self Compton radiation process.
{\mn Our  model implies} a  specific, recurring pattern in the
optical to gamma-ray emissions, made of high peaks and wide
troughs. Progressing accelerations {\mn caused} by spreading
reconnections {\mn will} produce an additional Synchrotron keV
component.
% during the optical rise.
Such outbursts  provide a diagnostics for enhanced acceleration of
electrons which can up-scatter photons into the
TeV range. %{\mn %We discuss applications
% Within the framework of our model
 We discuss how our model applies to the BL Lac object \pgp,
arguably the best candidate to now for
 high amplitude, recurring  modulations in its gamma-ray
emissions. We also consider other BL Lacs showing correlated keV -
TeV radiations
 such as Mrk 421.

\end{abstract}


%%%%%%%%%%%%%%%%%%%%%%%%%%%%%%%%%%%%%%%%%%%%%%%%%%%%%%%%%%%%%%%%%%%%
\section{Introduction}
%%%%%%%%%%%%%%%%%%%%%%%%%%%%%%%%%%%%%%%%%%%%%%%%%%%%%%%%%%%%%%%%%%%%

Blazars  are singled out among  the Active Galactic Nuclei (AGN)
by their relativistically {\af collimated }  jets with  bulk
Lorentz factors $\Gamma \sim 5 - 15$ (Urry \& Padovani 1995). They
are launched by a central super-massive black hole (SMBH) with
mass $M ~ 10^8 M_{\sun}$, and are Doppler-boosted when a jet
happens to be closely aligned with our line of sight.

{\af Blazars are extreme in several respects.  The observed
outputs are very bright, up to { $10^{49}$ erg/s } in isotropic
extrapolation, and strongly variable on diverse timescales from
years to minutes depending on the observed bands. They show highly
\emph{non-thermal} spectra with  energy distributions (SEDs)
constituted by two humps: one peaking in the $IR$-UV bands of
clear Synchrotron (S) origin; the other extending from hard X to
gamma rays, of likely inverse Compton (IC) nature (for basics see
Rybicki \& Lightman 1979).  Both are produced  by highly
relativistic electrons with random Lorentz factors up to $\gamma_p
\sim 10^3$ that inhabit the jets.

Two {main} Blazar flavors are  discerned (see Peterson 1997,
Ghisellini 2016). The BL Lac-type {sources}  feature two
comparable spectral humps.
 The Flat Spectrum Radio Quasars
(FSRQs), {on the other hand,} feature Compton-dominated spectra at
gamma-ray energies, {\mn but also}  a conspicuous Big Blue Bump
and the strong, broad optical emission lines common to many
quasars; { these thermal features yield evidence of a gas-rich
environment surrounding the SMBH out to some $10^{-1}$ pc.}
% These components are
%produced by \emph{thermal} electrons  in the {\aff central ring
%} of the accretion disk, and in a region at a distance $ r \sim
%R_{BLR} \simeq 3 \cdot 10^{17} $ cm, respectively {\af (see
%Peterson 2006)}.

% {\mt Accelerated particles within the relativistic
Relativistic electrons within the jet emit by synchrotron (S)
radiation} an observed (and isotropically extrapolated) luminosity
$L_S \propto n \, R^3 \, B^2 \, \gamma^2_p \,\, \Gamma ^4 \, ,$
%\be L_S \propto n \, R^3 \, B^2 \, \gamma^2_p \,\, \Gamma ^4 \, , \label{eq-1} \en
in terms  of the density $n$ of relativistic electrons
%with Lorentz factors up to $\gamma_p \sim 10^3 $
within the source size $R $,  and of the general magnetic field $B
\sim 1$ G threading the jet. {\af On the other hand,} the IC
  scattering operates on "seed"  photons on
conserving their number {\af while} upgrading their energy by
another factor $\gamma^2_p$, so as to yield  photon energies $h\,
\nu \sim h\, \nu_s \, \gamma^2_p$ and {related} luminosities $
L_{C} \propto L_{seed} \, R \, n \, \gamma_p^2  $.
%\be L_{IC} = L_{s} \, R \, n \, \gamma_p^2 %\propto B^2 \, \gamma_p^4
%\, \Gamma^4 . \label{eq-2} \en

The simplest source structure envisages as seeds the very S
photons that are radiated  by {\aff a single} population of
relativistic electrons in one, homogeneous zone of the jet, so as
to {\af yield } also a beamed flux  of gamma rays $L_{C} \sim L_S
\, n \, R\, \, \gamma_p^2$.
%\be L_{C} \sim L_S \, R\, \, \gamma_p^2.\en
Such a synchrotron self-Compton (SSC) radiation process (e.g.,
Maraschi, Ghisellini \& Celotti, 1992) is sufficient to account
also for the gamma-ray outputs in {tight } correlation with O - X
as featured by many BL Lacs in their {quiescent} states.

 %On the other hand,
 An interesting  development is constituted
 by  {\mn growing hints or signs of}  {\qq apparently recurrent }
  emissions in a number of  Blazars, BL Lacs in particular. On the origin
  and modeling of  these  phenomena  we will {\mn focus our attention}
 in the  present {\af paper.}

\section{A Binary Driver of Recurring Jet Instabilities }

% {Among  possible interpretations of their
% %  {quasi-periodic}
% {\qq repetitive} light curves listed by A15, we will concentrate
% on  {our conjecture that binary dynamics provides a common prime
% mover for all emissions}.  Any gravitationally bound binary system
% underlying \pg {ought to be currently engaged in } a slow-inspiral
% mode, that {\af periodically just \emph{perturbs} the main BH jet
% configuration while } it \emph{drives } magnetic reconnections
% (herafter, MRs), a process known to lead to {\aff effective } in
% situ electron accelerations as discussed later on.}

Interest in binary, massive BHs {\aff in AGNs started with
Begelman et al. 1980}, and is {\mt currently} being spurred by the
recent detections of {\aff bursts }of gravitational waves
constituting the event{\aff s} GW 150914 (Abbott et al. 2016a) and
{\aff GW 151226} (Abbott et al. 2016b). {\aff These} events have
been shown to originate from {binary systems} of  BHs with
intermediate masses $M \sim 50 \, M_{\sun}$. {\mgg The pairs }
have been caught in {\aff their} late inspiral stages, as they
{\aff {were accelerating }their orbital motions} under increasing
energy loss in gravitational waves to end up into catastrophic
coalescence as predicted. In fact, extensive numerical work
concerning binary BHs long pre-dated {\aff those
 events}, aimed at including   \emph{electromagnetic} outputs
from systems composed of two BHs and a warm disk or ring of gas
(see Palenzuela et al. 2010, and the review by Baumgarte \&
Shapiro 2011). Similar events are expected also in  binary systems
of two SMBHs (see Colpi  2014, Volonteri et al. 2015).

 {\aff Ever as}  such  emissions are dramatically
 enhanced during the final
coalescence,  they are {\aff numerically} found to set in {\af
gently} at much lower rates during the {\aff early,} long inspiral
stage. Similar computations can
%-- and ought to --
be focused as well on investigating the initially slow evolution
of {nearly} Keplerian orbits and their interactions with the {\mn
accretion} disk; we aim at understanding \emph{non-linear} light
curve shapes
% quasi-periodicities
recurring on scales of a few %  $\sim 1\, - 10$
years in the e.m. emissions,
  as {\af suggested by} current gamma-ray observations discussed in Sects. 3 and  4.

\begin{figure}
\vspace{-6cm}
   \centerline{\includegraphics[width=17cm, angle = 0]{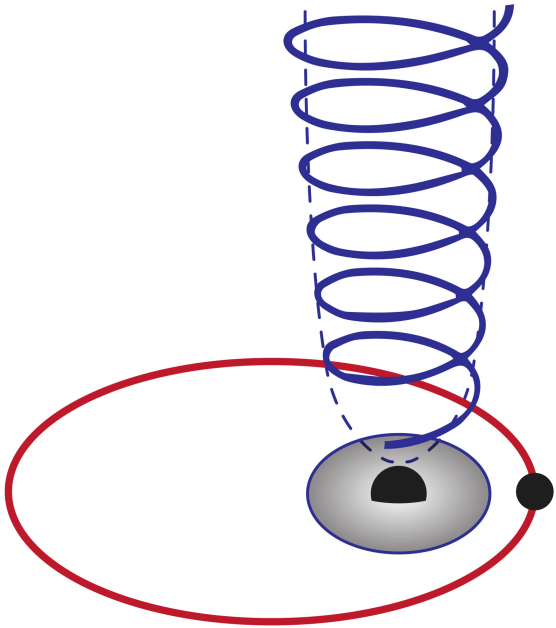}}
 \vspace{-4cm}
 \caption{{\mgg A schematic  of our model for the SMBH
binary  system underlying a BL Lac Blazar.  %\pgp.
%  in the frame of the primary black hole.
  The primary SMBH produces
  a relativistic jet towering well above its formation region in   the inner
  accretion disk (boundary marked by the thin curve). A SMBH companion
  orbits {around the system barycenter  close to the
  primary}, following the red orbit.
  The gravitational force  affects the jet
  base during its  passes close to the periastron, {\mn modulating its
  emissions to a pattern with high peaks followed by  long relaxations.}
  The orbital size is of order  $10^{16} \rm \, cm$ for a
  %{\qq
  total mass {\mn around} $\sim 10^8 \, M_{\odot}$,
  mass ratio $M_2/M_1 \sim 10^{-1}$, with an intrinsic  period $T \simeq 1.5 $~years.
  }}
\label{fig-0}
\end{figure}

% {\aff In fact}, several observational {hints} of periodicity have
% emerged in the recent literature, and involve {\af not only \pg
% %as {\aff {\mgg a} %the prime candidate} (see A15),
% but also } OJ 287 %as {\aff another strong case}
% (Valtonen et al. 2016), %adding to
% PKS 2155-304 (Sandrinelli et al. 2014)  {\qq and} a number of
% other, as yet weaker candidates including several BL Lacs and a
% few FSRQs as listed by Sandrinelli et al. (2014).

% Concerning \pg in particular, the extensive, multiple  analyses by
% A15 of the optical and  gamma-ray spikes and troughs observed
% during several years  have  produced an {interesting} case for
% quasi-periodicity  with the intrinsic period $T \simeq 1.5$ yr.
% {\af Our straight} explanation is in terms of

% {\alf No signs of period evolution have emerged yet, a
% circumstance that prompted us to entertain the simple
% interpretation given in Sect. 4.1}

%  The other BL Lac OJ~287 has been long
% analyzed (see  Valtonen et al. 2016) on using mainly optical data
% to notably   {bracket} the emission event at the end of 2015. It
% is to be stressed that in  such relatively close binaries the
% orbits  {may be }\emph{not} purely Keplerian and the events
% \emph{not} strictly periodic, as discussed  by Valtonen et al.
% (2016) for 0J 287 in particular.

{\af The} emissions of all Blazars are governed by the
hydrodynamic and magnetic {\aff properties} of the jet's
relativistic plasma outflow (see Sect. 1). Jet formation and
launching in many such sources appear to persist on gross average
over timescales of {\aff many } years. The considerable short term
variations
%{\mn - from {\aff the } mere detectability threshold to strong
%flares -}
observed in their radiative outputs have  {\qq often} been
considered to be fully erratic ({\aff see, e.g., Kelly et al.
2009)}. Jet formation and its {\af short term} instabilities are
then attributed to {\aff random variability  in the boundary
conditions occurring {\af at the base,} such as accretion rate;
alternatively,  disk instabilities {\mn (see Nixon \& King 2015,
and refs. therein) have been computed and discussed.  In the
present paper we consider an additional possibility.}
% However, the {\qq hints at}%  evidence for
% quasi-periodic emissions from \pg lead us to reconsider in such
% case the {\aff above} view.

We will investigate {\mn the} binary SMBH {scenario }{\aff
depicted in Fig. \ref{fig-0}, where the primary launches a main
jet that dominates the observed fluxes, but is perturbed  by the
secondary to  a considerable degree}, This we shall compute from
considering  the binary dynamics of a system with a total mass of
$\sim 10^8 \, M_{\odot}$}, mass ratio {\aff $M_2/M_1 \sim
10^{-1}$} and a current orbital size around several milli-pc.
% , that ought to yield next maxima in early 2017 and in 2019.
{\af Specifically,} the bulk properties of the main jet are
affected by episodes of {\alf shaking}, squeezing and twisting
caused by differential gravitational and electromagnetic stresses
along the non-circular, or even {not simply} Keplerian orbits. The
results include time-modulations of the average \emph{strength}
$B(t)$ of the magnetic field, coordinated  with changes in the
local \emph{topology} of {\mn B}-lines by reconnecting and
rearranging those that angularly diverge across narrow layers, as
we shall discuss in the next Section.

Here we just note that  passive geometrical effects of gentle beam
redirection along steadily curved or helical jets as advocated,
e.g., by Raiteri et al. (2015) and {\mmm discussed } by A15 may be
{\aff re-interpreted } in terms of initially mild {\mn and slow}
dynamical effects that take place during the quasi-period. {\mmm
Our discussion} will focus instead onto the  dynamical
\emph{driving} in the jet of sites suitable for magnetic
reconnections to take place.

\section{Unstable Jets}

{\mn We consider how the progression from large-scale jet
instabilities proceeds through an intermediate stage of tearing
perturbations to end up into kinetic effects that spread out
electron acceleration to high energies.}

We {\af base our discussion on} the canonical  SSC radiation
process recalled in Sect. 1. As to  the primary S emission,  we
adopt standard values of the average magnetic field $ B \sim
10^{-1} - 1$ G in the source frame, at the height of some $
10^{17}$ cm above the primary SMBH. These are modulated in time
% as discussed above,
by factors of about three on timescales of  {\mn a few years }by
compressing and bending  {\mn B} {\mt as discussed below}.

The other basic component for  S emission is constituted by highly
relativistic electrons. {\mtg Their} acceleration  occurs in
regions within the jet {\af that are} affected by magnetic field
lines {packed, sheared and reconnected} so as to
induce macroscopic and kinetic effects.  % \footnote{
For a review covering the magnetic reconnection  (MR)  theory and
{\mmmm the ongoing} numerical simulations see Kagan et al. (2015),
also Melzani et al. (2014). For recent theoretical developments on
reconnecting structures in collisionless plasmas, see Coppi
(2016); for detailed observational results concerning
reconnections in a magnetospheric environment, see Burch et al.
(2016).
% simulations addressing  mass ratios approaching
% the proton-electron value).

The intermediate output of MRs and  {\aff associated tearing % mode
instabilities } is constituted by strings of  "magnetic islands"
or plasmoids, many of which then {\af merge into a  few giant
ones}. Meanwhile, {\hh on} the \emph{kinetic} side  {\hh sharply
sheared, or even} annihilating \textbf{B}-lines induce strong if
confined \textbf{E}-lines; these {\hh repeatedly and efficiently}
accelerate electrons {\textit{in situ}} to high values of
 $\gamma \sim 10^3$, particularly around
the {\af giant}, coalescing  plasmoids %{\mn that tend to coalesce}
and in the intervening gaps. {\mmmm Next,} we  focus on the  {\af
key parameters governing these two, related} sides of the {\alf
processes} {\mmmm occurring in magnetized jets}.

%The reconnections  are known (see Sironi \& Spitkovsky 2014,
%{\aff and the review by Kagan et al. 12015}) to take place
%through excitation of tearing modes that fragment the plasma into
%"magnetic island"  some of which then merge into giant plasmoids;
%around and between them the annihilations of opposite {\aff B}
%lines induce strong if localized {\aff E} fields that
%efficiently \emph{accelerate} electrons to highly relativistic
%energies. As we will see in the next Section, the
%governing,{\aff average} magnetization parameter $\sigma_j$ is
%then considerably increased, and even more is the similar local
%$\sigma_e$ parameter governing the electron accelerations.

%\end{document}

\subsection{The macroscopic side}

The relevant parameter for stability on large {\mmmm scales} is
provided by the {\mgg  \emph{bulk} {magnetization} of the jet}
 \be \sigma_j = \frac{\bar{B}^2}{4\, \pi\, n_p \, m_p\, c^2\, \Gamma^2} \, . \label{eq-1} \en
 This depends on the average  field {that appears}
in the {magnetic stress } $\bar{B^2}/4\, \pi$, and  on the kinetic
stress   $n\, m_p\, c^2\, \Gamma^2$ % due mainly to
{\mmm dominated by} the jet protons
%{\mgg  $\Gamma$ {\mmm being} the bulk Lorentz factor}
 (see Celotti \& Ghisellini 2008).
%and the bulk Lorentz factor $\Gamma$.

The condition $\sigma_j \lesssim 1$ {yields  stability} of a jet
on large scales, for example, on the scales $
 10^{16} - 10^{17}$ cm numerically investigated by, e.g., Mignone et al.
 (2013); {\mmm on the other hand, {\hh values} $\sigma_j \gtrsim 1$ {\hh are} found by the same
 authors to promote jet instabilities}.
Thus the magnetization $\sigma_j$ constitutes the main
\emph{parameter} governing the {overall} stability of a jet
configuration on large scales.

% and $\lesssim 10^{-1}$ to hold for the {\aff former} and for the
% {\aff latter}, respectively. So the former are expected to host
%{\mgg Jets with $\sigma_j \gtrsim 1$

%\end{document}

Using values of  $B \sim (0.1 - 1) $ G,   $n \sim 10^{-1} - 10^2
\, \rm cm^{-3} $ related to the accretion rates on the primary
SMBH, and values $\Gamma \sim 10$, as derived, e.g., by Paggi et
al. (2009),   Tavani et al. (2015), and observed by Hovatta et al.
(2009), we find $\sigma_j < 10^{-1}$ to hold in FSRQs, while
$\sigma_j \lesssim 1$ prevails in BL Lacs. Thus the latter appear
to  sit on the brink of instability, and may be driven
\emph{unstable} by  minor dynamical perturbations such as may be
induced by a binary BH companion.  On evaluating at $T/10$
(details are given in the Appendix) the period fraction with
strong primary acceleration while revolving  around the
barycenter, it is seen that  the jet  cannot remain connected over
scales $\ell > c\, T/10 \sim 10^{17}$ cm against dynamical
perturbations that shake  its base on  an effective timescale of a
few months. This also sets an upper limit to any spread along the
axis of the sources of correlated emissions.

%\end{document}

\subsection{The kinetic side}

At the opposite extreme constituted by the  local \emph{ kinetic}
level, acceleration of electrons to high energies is governed by
repeated MR events.
% These are associated with
% plasmoid strings formed by  tearing modes that subsequently merge
% favoring the formation of large electric fields  } (e.g., Kagan et
% al. 2015).
The acceleration process has been {\af parametrized } and
numerically computed in terms of the {\mgg electron \emph{local}
 {magnetization}}
%of "cold" electrons
(see the review of Kagan et al. 2015)
% Sironi \& Spitkovsky 2014)
\be \sigma_e = \frac{B^2}{4\, \pi\, %} {\cal E}_e}
 n_e \, m_e\, c^2} ,
 \label{eq-2} \en
which depends on the \emph{local} magnetic field $B$ and electron
density $n_e = n_p$, given their mass $m_e$.

Many  numerical calculations have been carried out in the range of
$\sigma_e$ {\af values } from tens to up to several hundreds (see
Sironi \& Spitkovsky 2014, Melzani et al. 2014). Remarkably, for
increasing $\sigma_e  >  10^2 $  the resulting electron energy
distributions approach  flat power-laws $n(\gamma)  \propto
\gamma^{-1.5}$ up to, or asymptotically exceeding $\gamma \sim
10^3$ {\aff  limited by computing times} (e.g., Melzani et al.
2014). % Note
{\mtg We stress  that}
% {\mgg  for comparable local and average {strengths} of the magnetic field,}
in the simplest case the relation
 \be \sigma_e \gg \frac{m_p}{m_e} \,
% \Gamma^2 \,
\sigma_j \sim  10^3 \, \sigma_j \label{eq-3} \en holds, so that
the electron energies can grow to values $\gamma \sim 10^3$ and
{\aff possibly } beyond
  % \rm \, a \, few \, 10^3$.}}
 as soon as bulk instability
conditions are established by values\footnote{{\mmm Note that Eq.
\ref{eq-3} as it stands applies for values of $\gamma \lesssim
m_p/m_e$. For larger values, the electron contribution to the
total kinetic energy should not be neglected; $\sigma_j$ is
correspondingly decreased, and so is the effective
% electron magnetization
 $\sigma_e$. This feature may imply
self-regulation {\alf to} limiting energies
 $\gamma \sim 10^4$.}}
 of $\sigma_j > 1$.
 In other words, {\mgg conditions conducive to} bulk instability act as a
\emph{trigger } for strong particle accelerations {\alf to occur}.
Note, however, that the electron magnetization in  Eq. 2
%\ref{eq-3}
affects the particle energy distribution function only in the
presence of \emph{topological} configurations of the local
magnetic field {\mmmm that include {\hh sheared layers or} close
reversals}
 {\aff conducive to} {\mmm tearing and} reconnections in the
  jet's collisionless  plasma.
 % \footnote{For a recent measurement of  collisionless reconnection in
 %a magnetospheric environment, see Burch et al. (2016).}.

%\end{document}

Extensive and detailed computations have recently been presented
by Yuan et al. 2016 concerning the link between macroscopic field
configurations with higher magnetization and kinetic conditions
conducive to acceleration. We conclude that a slow \emph{time}
shift of macroscopic conditions toward increased $\sigma_j$ and
sheared \textbf{B} topology can \emph{end up } into triggering and
enhancing local accelerations governed by high values of
$\sigma_e$.

%\end{document}

\subsection{The binary drive}

Thus current knowledge {\mn  backs our expectation} that {\mn in
BL Lac jets -- by themselves in near-critical conditions -- even
modest perturbations that increase B can easily drive the
structure beyond the threshold of macroscopic instabilities. Soon
after, the  conditions will progress toward strong kinetic
instabilities if  the macroscopic modes induce local torsion and
shear of \textbf{B}}. The ensuing  scale for field shearing {\mmmm
will be} close to the longitudinal mode wavelengths experienced by
the jet, that are of order of $10^{16} - 10^{17}\, $cm  as {\mmm
numerically computed}, e.g., by Mignone et al. 2013 (see also
Striani et al. 2016).

%\end{document}

In the case of a binary  SMBH system, we  envision as
\emph{driving} agent the dynamical perturbations of the magnetic
jet configuration  by the  companion orbiting  around
 the accretion disk, from which the main jet is launched.
When the perturbations  {\aff occur}, the field  \textbf{B}  not
only is locally \emph{compressed} to larger strength $B + \delta B
\simeq \,$ a few times $\, B$, but also is locally \emph{sheared }
so that regions of {\mmm sharply different} magnetic field lines
are  drawn together.

In a nutshell, binary perturbations \emph{drive} macroscopic jet
instabilities that offer many suitable locations for MR to occur
and \emph{trigger}  effective electron acceleration; the ensuing
emissions are detailed next.
%\end{document}

\section{Emissions and Their Interpretations}

{\mn The basic feature to be expected from the binary drive of jet
instabilities as discussed in Sect. 3 concerns a recurring,
non-linear spike--trough pattern in the  light curves, reflecting
the stroke--relaxation nature of the dynamical drive along a
Keplerian  elliptical orbit.}

\subsection{Light curves}

{\mn To be specific, we shall refer} to the BL Lac object \pg as
an interesting candidate for recurrent non-linear gamma-ray
emission as observed and discussed by Ackermann et al. (2015,
henceforth A15). The source features  {\mn in its}
\emph{continuously} monitored gamma-ray emission high amplitude
modulations that recur with an apparent period $P \simeq 2.2$ yr;
at  the source redshift $z \simeq 0.5$ (see Danforth et al. 2010)
this corresponds to an intrinsic value $T \simeq 1.5$ yr. {\mn
Three full (in fact, 3.5), consecutive periods are continuously
covered by the existing {\mnn gamma-ray} % data
observations
 {\mnn of \pg
presented in A15; these data % substantially
extend the dynamic range (number of cycles and % modulation features
amplitudes) of previous blazar periodicity analyses such as that
of Vaughan et al. 2016. From the A15 analysis of \pgp,} a next
gamma-ray peak is expected in early 2017.

%-------

 \begin{figure}
% \vspace*{-4.5cm}
%  \centerline{\includegraphics[width=15cm, angle = 0]{fig3-plus.jpg}}
%
 \centerline{\includegraphics[width=18cm,angle =-0]{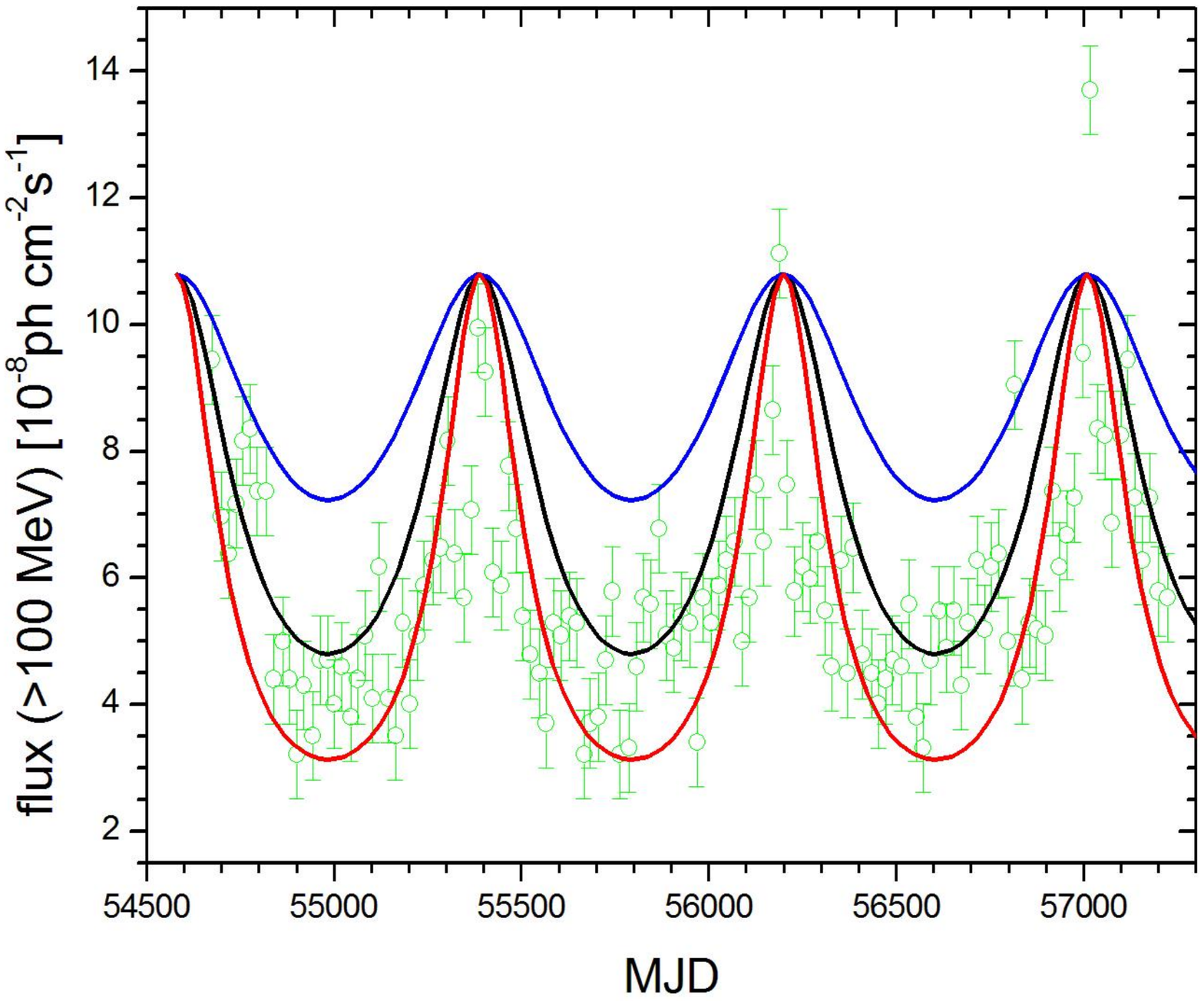} }  %{fig3nn.jpg}}
% \vspace*{-3.5cm}
     \caption{The gamma-ray light curve of \pg (green data points and error bars from A15)
     is compared with
      the predictions from our model, in which the gamma-ray flux is proportional to the gravitational
      force after $f (t)\propto F(r)/F(r_{max})$ (cf. Eq. \ref{eq-4} in the Appendix) along the elliptical orbit at a distance $r$ from the focus.
      We used the apparent period  $P = 2.2$ yr (see Sect 4.2), and  three values
      for the eccentricity:  $\epsilon=0.1$ (blue curve),
      $\epsilon=0.2$ (black  curve) and $\epsilon=0.3$ (red curve).  For $\epsilon=0$ the orbit would be
      circular and the light curve just flat,  apart from minor random fluctuations.}
 \label{fig-app-1}
 \end{figure}

{\mn In the Appendix we derive in detail a simple model for the
gamma-ray light curve based on the gravitational  modulation in a
binary system, as impressed by the companion on the primary BH and
its associated jet. Our result is  presented here in Fig.}
\ref{fig-app-1}, {\mn along with the 7-year data concerning \pg
from A15}. {\mn It is seen that our model yields  a
\emph{non-linear} pattern of the light curve with prominent peaks
and wide troughs, that for an eccentricity $\epsilon \simeq 0.2$
 is quantitatively very close to the overlaid data points.}

\subsection{Spectral distributions}

The basic {\mn spectral change} expected from  compressed and
sheared magnetic fields in the jet is twofold. {\mn To begin
with,} we expect an increased \emph{overall} {\af SSC emissivity},
{\mn led by  the optical S emission and followed {\mn after
 some weeks} by IC radiation} at GeV energies.
%
% as {suggested} by A15 {\mn in case of \pg} and by Cohen et al.
% (2014) {to occur }{\mta along  the rising branches of the optical
% emission of} Fig. \ref{fig-1}; the process may {take up to} {some
% } weeks.
%
Such a direct correlation is well compatible with the data as
discussed by A15,  {\mn but is also expected to be appreciably
smeared out along the jet axis by  outflow and long cooling times
for both  component sources (see Table 1 and its caption).}

% {\mn Table 1 shows the  parameters that we used to model the
% spectrum of the BL Lac system \pg during an episode of enhanced
% keV emission together with opti

%cal and TeV emission near MJD

% 56000}.
%\end{document}

In  detail, the   implied
%of increased bulk
%magnetization $\sigma_j$ in the jet is
% an  increased magnetic field $B$ implying
 enhancements of  S, {\mmmm optical} radiation and of its IC
   counterpart at GeV energies
are  shown by the spectra marked by red color in Fig. 3 with
overlaid data.
%~\ref{fig-3}.
Such flux variations in  both the R-band and the GeV component
range by a factor {\hh around} 3 from minima to maxima.
%
% in  Fig.~\ref{fig-1} the feature {is apparent} at phases near
% JD~55300 and MJD~56000.
In conclusion, the emissions in the optical, the soft X and  the
GeV bands  appear to  generally correlate along the spectra in
shape and timing  in the way described by the canonical SSC
radiation model.

On the other hand, the behavior at keV energies of \pg appears to
{\mn stray outside } the above framework, and has been reported as
"enigmatic" (see, e.g.,  Raiteri et al. 2915). For one, the
increases during episodes of optical rise differ in softer X-ray
ranges: they rise from a factor of 3  in the range 0.3 - 2 keV,
but are  up to factors 5 - 10 in the  2-10 keV range as reported
by Aleksic et al. (2015) for the episode around MJD 56000. For
two, other BL Lacs such as Mkn 421 appear to share with \pg such
keV features as missing correlation  with neighboring X rays, and
positive correlations with TeV emissions (Balokovic et al., 2016).

%On the other hand, a  second change is expected from additional,
%spreading electron
%accelerations is best seen in the SED. %{\mmm triggered},

%In fact,  this will imply an increased  upper energy {\mmm break }
%$\gamma_b$,   associated with  X-ray emission at keV energies
%{\mgg at and } beyond the {\aff steady}, the first hump in the
%SED, as in fact discerned  in Fig. \ref{fig-3}, see also Table 1.

%---

\subsection{An additional spectral component in X rays }

%On the other hand,  the \emph{additional}  X-ray emission
% {\aff above 3 \lesssimkeV}
%is {not} straightforwardly correlated with the O - gamma-ray
%ranges, a {\mmmm telltale feature}  toward understanding {\mn BL
%Lac X-ray} properties % of \pg and {possibly of other sources
%(such as \pgp, A15, and Mrk 421, Balokovic et al., 2016).
%.......
%{\mn In the case of \pgp,}
%{\mtg we % note

%\end{document}

To interpret such a behavior,  we recall the contents of Sect. 3.
In particular, in Sect. 3.1 we noted that a  BL Lac jet is
normally  metastable, so that a  limited increase  of $B$ can
drive it first toward  macroscopic,
 and then into kinetic instability conditions.

 In a first stage  the  near-critical
bulk magnetization $\sigma_j \lesssim ~ 1$ is increased by a
factor of a few, as deduced from the optical light curves {\alf
enhanced during their
100-day rise}. %(see Fig.\ref{fig-1}).
This implies a drift from marginal values of $\sigma_j$  % \lesssim 1$
into the fully critical {range} for macroscopic instabilities. If
these produce   twisted and inverted $ \bf{B}$ lines
 %proportional increase of $\sigma_e$ ,
a second stage arises, driven by increasing local magnetizations
$\sigma_e \propto \sigma_j$ after Eq. 3.
% even more important, being related in time, and compellingly
% connected with our theoretical picture,
These  enhance  MRs and cause extensive particle acceleration.
Based on current numerical simulations that are exploring
different regimes of ion-electron plasmas (see Melzani et al.
2014, Kogan et al. 2015), we derive
 that the maximal electron energies
 %depending on $\sigma_e$
 can approach  values  $\gamma_p \sim 10^4 $ for values of $\sigma_e$
exceeding $10^2$. Such  enhanced accelerations will \emph{ add } a
 stronger and harder spectral component in the X-ray range  {\alf
around and above $1$ keV}, as indeed detected in at least two
episodes {\af within} the currently available data concerning  \pg
(A15).
% (see Fig.~\ref{fig-1}, middle panel).
% Note how such enhancements are started during the
% \emph{rise } of the optical light curve \mmmm near MJD 56000 and
% 57000.
%
% We stress  that s
The  X-ray flux increase
%{\aff related } to the rising branch of the optical emission
 amounts  {\mtt to}   factors
of 5-10, i.e., substantially \emph{larger} than the related rise
of optical and GeV emissions. Clearly,  here one-zone, single
population modelling of the X-ray emission together with the
optical and GeV emissions would not be adequate. Our picture, on
the other hand,   implies  additional particle acceleration {\mmmm
triggered} in nearby zones with {increasing } values of
$\sigma_e$. The corresponding SEDs for one of such  episodes
 are presented in Fig. 3.

%%%%%~\ref{fig-3}.

%\end{document}

\begin{table}[!t]
\begin{center}
\centerline{\bf Table 1: parameters for modelling the spectral
states of PG 1553+113.} \vspace*{.2cm}
 \small \noindent
\begin{tabular}{|c|c|c|c|c|c|c|c|c|c|}
  \hline
  \bf{State} & \bf{Comp.} & $\bf{\Gamma}$ & \bf{B(G)} & \bf{R(cm)} & $\bf{K(cm^{-3})}$ &
  $\bf{\gamma_b}$ & $\bf{\gamma_{min}}$ & $\bf{\alpha_1}$ & $\bf{\alpha_2}$ \\ % & $\bf{\tau_{c}/\tau_{cr}}$ \\
  \hline
  low &  gray & 15 & 0.25   & $4.0\cdot 10^{16}$ & 0.6 & $9\cdot 10^3$ &  $4\cdot 10^3$ & 2 &3.9 \\ % & 1.3 \\
%          & c II & 15 & 0.3 & $3.5\times10^{16}$ & 1.4 & $7000$ & $3500$ &2&5 \\ % & 1.0 \\
%          &single& 15 & 0.3 & $3.5\times10^{16}$  & 1.8 & 7000 & 60 & 1.8 &5 \\ % &1.0\\

  \hline
enhanced soft & red & 15 & 0.29   & $4.0\cdot 10^{16}$ & 0.6 & $9\cdot 10^3$ &  $3\cdot 10^3$ &2&3.9 \\ % &1.2 \\
enhanced hard    & blue dashed& 15 & 0.29 & $3.5\cdot 10^{16}$ & 0.1 & $3\cdot 10^4$ & $5\cdot 10^3$ &2&4.3 \\ % &1.1 \\
%         &single& 15 & 0.3 & $3.5\times10^{16}$  & 1.8 & 6500 & 30 & 1.8 & 5& \\ % 1.1 \\
  \hline
\end{tabular}
\end{center}
{\small \mmm The Table summarizes the parameters used in our SSC
spectral modelling of \pg  ({\mn see Fig. 3.)} The Table provides
values for the bulk Lorentz factor $\Gamma$, the average magnetic
field $B$ inducing Synchrotron emission, the source size $R$,  the
electron energy distribution $n(\gamma) = K \,\gamma_b^{-1}\,
(\gamma/\gamma_b)^{-\alpha}$ with} normalization K,  upper break
at $\gamma_b$  and {\mn  lower bound} $\gamma_{min}$, {\mn and}
power-law indices $\alpha_1$ below and $\alpha_2$ above  the
break. The associated cooling times for S and for IC (see Rybicki
\& Lightman 1979) read $t_c \simeq 2 \, R/c \simeq 30$ d .
%  and the {\alf ratio of the cooling} $\tau_c$ to the
% crossing time $\tau_{cr}$.
 \vspace*{.5cm}
\end{table}

% \end{document}

%These become dramatic on approaching the short  final plunge as
%impressively illustrated by Panenzuela et al. 2010. On the other
%hand, during the \emph{long } inspiral stage they will cause
%gentle distortions and squeezing  of a jet and of the associated
%magnetic field lines eventually leading to reconnection
%instability; we note that the passive geometrical effects of beam
%re-direction along steadily curved or helical jets as advocated,
%e.g., by Raiteri et al. 2011 may be re-interpreted in terms of
%initially mild dynamical effects that take place during the
%quasi-period. Primarily, however, investigations focused onto  the
%detailed dynamical \emph{triggering} of magnetic reconnections
%regions are badly wanted.

%\end{document}

 \begin{figure}
%\vspace*{-4cm}
%  \centerline{\includegraphics[width=15cm, angle = 0]{fig3-plus.jpg}}
%
  \centerline{\includegraphics[width=19cm,angle =0]{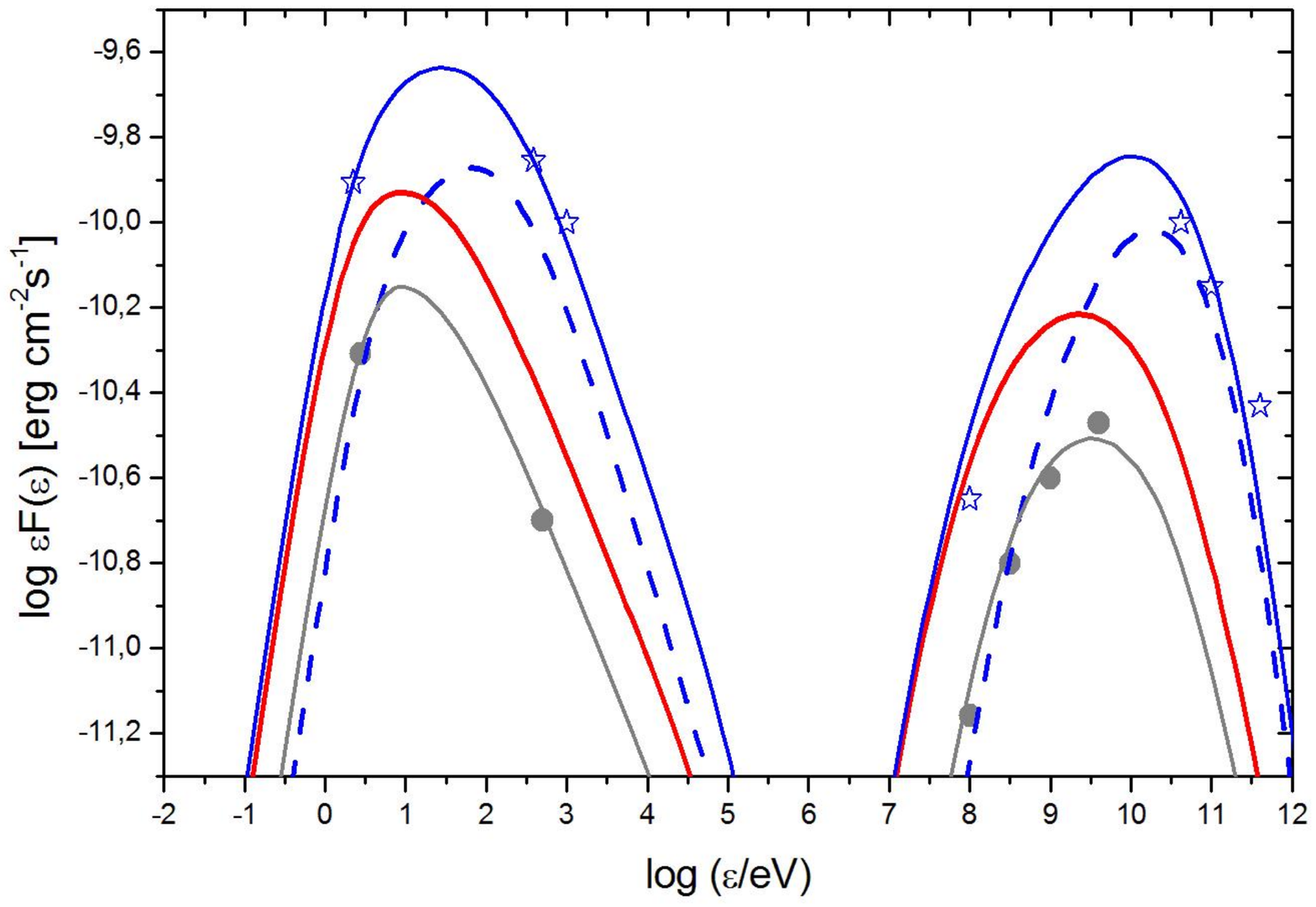} }  %{fig3nn.jpg}}
 % \vspace*{-7.5cm}
     \caption{ Spectral energy distribution (SED) of the
     emissions by electrons accelerated in the  jet  of the BL Lac
    \pg. Spectral data are from Ackermann et al. (2015) and
    Aleksic et al. (2015).
The gray line shows the spectral state
 during {\mn normal} conditions.
The {\mn blue, dashed  curve} represents the \emph{harder},
increased  Synchrotron and IC  components
% near the top  of the light curves observed around MJD ...
%\emph{additional },  harder component
resulting from the extra {\mn electron} acceleration discussed in
Sect. 4.3. The total emission in the \emph{enhanced } state is
represented by the {\mn blue } solid curve. {\hh EBL de-absorption
in the TeV  data after Aleksic et al. 2012}. }

%\end{document}

 \label{fig-3}
 \end{figure}

\section{Discussion and Conclusions}

% {We have focused on the {\alf cyclic } pattern  observed in the
% light curves from optical frequencies to gamma-ray energies in the
% BL Lac-type Blazar \pgp, which shows %  a basal random flickering
%  recurring peaks and troughs % some $3-5$ times  {higher }
%  with an apparent period $P \simeq 2.18$ yr. We have  entertained a {\hh
% simple} interpretation based on a binary nature of the underlying
% SMBH, and explored its implications.}
%
%  throughout four well
% developed and extensively computed domains of astrophysics: binary
% Black Holes, large-scale MHD instabilities of jets, small-scale
% magnetic reconnections in the jet's thin plasma that energize
% electrons, the ensuing SSC radiations. }

 {\alf We have argued in Sect. 2 that}
dynamic {events} in Blazar jets  - and specifically in BL Lac
Objects - originate from an  underlying \emph{binary } system of
two Super Massive Black Holes with total mass $M \sim 10^8 \,
M_{\odot}$. Binary dynamics constitutes, as we stated  in Sect. 2
and explained in Sect. 3, the \emph{prime mover} for a chain of
events in the jet: it recurrently drives bulk plasma instabilities
in the metastable magnetized jet launched by the primary;  in
turn, these trigger local tearing instabilities and \textit{in
situ} electron acceleration to high energies corresponding to, and
at times exceeding $\gamma \sim 10^3$.

{\mn Such local instabilities and accelerations are known to
%are  theoretically and numerically known to
arise in narrow layers of a collisionless plasma, where
\emph{reconnections} occur among adjacent ${\mn B}$-lines that
%\emph{reconnections}
are strongly sheared or even reversed}, as we detailed in Sects.
3.1 and 3.2.  Such processes are bound to be triggered during
dynamic episodes of large-scale squeezing and shearing {\alf of a
magnetized jet.} A conducive context for  such episodes {\alf to
occur} is provided by \emph{binary} SMBH dynamics (as discussed in
Sect. 3.3), that also provides the simplest interpretation  for
the   {\qq recurrent}, non-linear pattern of peaks and trough  in
the light curves of the BL Lac source \pg as discussed in Sect.
4.1.

{ We have  analyzed in Sect. 4.2 the different \emph{spectral}
states that arise in this source  and  exhibit  {\alf repetition}.
They start with a recurrent   increase of the optical Synchrotron
and of the inverse Compton  GeV radiations produced by the same
population of energetic electrons. We {\alf have  stressed} in
Sect. 4.3 the \emph{additional } spectral component that  {\qq has
been observed} in the keV range as the optical light curve
approaches a peak. We {\alf traced} it back to \emph{spreading}
accelerations triggered by additional reconnections stimulated by
binary dynamics.  We expect this component  {\alf to correlate}
with TeV enhancement, as occurring in other BL Lacs.

So we conclude that {at times} \emph{two } spectral states
 overlap {\alf in the emissions from } \pg. One is
well represented as quasi-periodic {\alf emission}, that envisages
the optical - soft X-ray components { and the  GeV  photons as
linked components of the same SSC radiation process;}  the latter
therefore should correlate with the former, after a lag {of some
weeks}. The {\alf other},  {more} energetically demanding state
requires \emph{additional} and {higher } electron accelerations
that produce a harder X-ray peak, {as a component (see Section
4.3) observed in two out of three monitored
instances. % (see Fig. 3, blue curve);
 this we expect to correlate} with TeV enhancements, not
{necessarily} with the optical --  GeV radiations.

Next we summarize our overall  conclusions, and specifically two
points. First, we have described how  a BL Lac jet, normally
sitting on the brink of instability, can undergo recurrent
\emph{stresses} from the binary dynamics sufficient to stimulate
cyclic emissions. In fact, increasing bulk magnetization leads to
large scale instabilities that cascade down to small scale  MRs
and ultimately percolate to the kinetic level; there they
accelerate the electrons that emit from the optical to the
gamma-ray band,
 and sometimes beyond.

 Second,  the binary Keplerian motion drives strokes and allows relaxations corresponding to orbital
 arcs close to, and far away from the periastron, respectively. So it imprints a specific
 non-linear{\qq \emph{pattern} of sharp peaks and wide troughs in} the light curves at different
 frequencies; these are particularly sharp in gamma rays that have been
 continuously monitored from space. Such a pattern  can provide diagnostics
 for orbital parameters,  in particular for the orbital eccentricity
  $\epsilon$ as detailed in the Appendix,  and
 used in Fig. 2 to derive for \pg a value $\varepsilon \simeq  0.2$.

As to predictions and prospects,  in the framework of the next
gamma-ray peak expected  for PG~1553 in early 2017 by A15, we
stress the following points. \textit{(1)} The new peak should
appear in reasonable phase {\qq agreement (within the time
binning) with the previous three occurrences. \textit{(2)} Thus
the optical emission should be monitored as continuously as
possible across the peak phase range,  with the help  of
telescopes strategically coordinated. The optical emission may be
quite structured as shown by Figs. 1 and 2  of A15; in our picture
we expect the main optical increase to begin {several days to some
weeks } before the smoother and better sampled gamma-ray peak,
which can {sound a last call (as it were)} for aimed X-ray
observations. \textit{(3)} In fact, the X-ray extra component in
the keV range is expected to be triggered
% stimulated
by expanding reconnections that occur just at, or  on the approach
to the periastron. Current data have the peak of additional X-ray
emission  corresponding in one case to the gamma-ray peak near MJD
57000 (to within the time binning); in another case,  preceding he
gamma-ray peak by $\sim$100 days. Our model contemplates enhanced
keV  emission somewhere between the rise of the optical and of the
gamma-ray light curves, on account of the spread along the jet of
the S and the IC  main sources implied by their long cooling times
(see Sect. 4.3, and Table 1).

%\end{document}

%\end{document}

It is interesting to recall that a similar radiative behavior to
\pg is shared by other BL Lacs, in particular by Mrk 421 with its
long puzzling, X-ray outbursts correlated with TeV activity but
not with optical and GeV emissions (se Raiteri et al. 2015,
Balokovic et al. 2015). Some of these sources (listed, e.g.,  by
Sandrinelli et al. 2014) have been proposed also as candidates for
recurring behavior in the optical and gamma-ray bands.

%In addition, among the FSRQs {\alf monitored over long stretches},
%3C 454.3 in particular yields evidence of related {emission}
%patterns comprising gentle S modulations over several years, with
%outbursts of gamma-ray flares setting in close to their maxima
%(e.g., Giommi 2015). {On the other hand, in such FSRQs the GeV
%range features strong variability on scales of minutes (Ackermann
%et al. 2016) that calls for specific source structures as  we will
%present elsewhere.}

%\end{document}

We note that the %current
\emph{dominance} of
%{\qq claimed} or {\mmmmn suspected }
possible BL Lacs' periodicities over FSRQs' in the set of current
%  {acceptable}
candidates may just reflect selection favoring the {former } on
account of their generally {smaller redshifts  and  lack of
disturbing $O - UV$ background from BLs and BBB pointed out in
Sect. 1}. {On the other hand, intrinsic association of BL Lacs
with binary sources is suggested by a number of  circumstances}.
First, as a consequence of hierarchical galaxy formation, the BL
Lacs as {aged} AGNs (Cavaliere \& D'Elia 2002, B\"ottcher \&
Dermer 2002) must have witnessed several merging events {\mmmmn
involving} {their host} nuclei. Thus, expanding on the pioneering
suggestion by Begelman et al. (1980) such nuclei would be more
massive and likely to have accrued several SMBHs to {eventually}
form a dominant \emph{binary } pair. In addition, the intrinsic
meta-stability of BL Lac jets (see Sect. 3) makes the latter
susceptible even to minor dynamic disturbances in forming a
{recurrent} binary-driven source. {Furthermore, the absence in BL
Lac spectra of thermal features as recalled above points toward
gas-poor conditions  in the vicinity of the central SMBH binary;
these prevent local energy dissipations from occurring,  and allow
for longer binary lifetimes implying higher intrinsic statistics
(see Colpi 2914 and refs. therein), limited mainly by e.m. losses,
and eventually cut off by a gravitational catastrophe (see Sect.
2).}

Finally,  we consider the  A15 prediction concerning a next
maximum of \pg {\qq in early 2017, with a spread of some 20-40
days due to time binning}; this would mark an obvious  plus for
the periodic view. On the quantitative side, extending the
sequence of fully observed cycles from 3 to 4
 would considerably  strengthen the confidence level
 $0.99$ recalled in Sect. 4. Thus by early 2017 one had better be \emph{prepared} with
astrophysical analysis of the currently known cycles as undertaken
here, and with coordinated  observations of keV -- TeV emissions.
We add that evidence of any secondary jet would provide a
tell-tale  signature of binarity that deserves a close search,
despite a number of difficulties: the expected weak power ratio
$\propto M_2/M_1 \sim 10^{- 1}$ even in the favorable case of both
jets fed by accretion at a common fraction of their Eddington
rates; the uncertainty about the jet relative angle; and the range
in the time phase caused  by  precessional shifts.
 %{(on scales of several times P)} the phase of the secondary
%contribution along the main light curve. }

{\qq  These tasks are demanding, but the} reward of
% any strong
confirmed evidence for binarity would be worth of keen
observational efforts. It is genarally agree that understanding
even a single SMBH binary would open a new window on the vexing
issue of co-evolution of galaxies and their central BHs (see Lapi
et al. 2014, Volonteri et al. 2015). In addition, a few confirmed
SMBH binaries with assessed orbital  parameters such as proposed
here may be of keen value for planning searches of giant,
 slow bursts of gravitational waves with space interferometers
such as eLISA  (see eLISA Consortium, 2013). In fact, the path
from host galaxy mergers to a sub-pc binary doomed to eventually
coalesce, is complex (see Colpi 2015) and currently still fraught
with a number of uncertain steps ("stalling" in particular),  that
not even massive numerical computations or simulations have been
able to fix to now.

%\end{document}

% Non-dynamical  mechanisms listed in A15 are equally speculative as
% binarity, but  (at variance with  stimulated reconnections)
% unlikely to lead to organized accelerations, besides being  of
% lesser impact on astrophysics at large.

% In perspective, in the light of increasing evidence for cyclic
% behaviors and possibly a binary nature of  a subset of Blazars,
% {\alf we expect a new} {\alf rich} chapter to open open up in the
% saga of AGNs. Future research will  extract from simulations,
% magnetohydrodynamical calculations and joint  kinetic computations
% the detailed sequence of events that produce in the jets magnetic
% twisting, shearing and reconnections under the drive of a binary
% companion.

\vskip .4cm \acknowledgements {\hh We are indebted to Bruno Coppi
for enlightening discussions on magnetic field reconnecting
structures in collisionless plasmas.} { We thank  our referee for
recommending a generally conservative attitude towards {\mn the
quasi-periodicity of \pgp), and for pressing us to specify the
role of light curve profiles, as we did in the revised MS}.
% and of direct clues of binarity; finally, for
%inducing us to  spell  out  in Sect.  5 purpose and timeliness of
%our paper in the framework of the predicted next period of PG 1553
%in early 2017.}
Work partially supported by the ASI grant no. I/028/12/2.

\newpage

%\end{document}

\section*{References}

%\begin{thebibliography}{}

%\bibitem[]{}

\noindent Abbott et al., 2016a, Phys. Rev. Letters, 116, 061102

\noindent Abbott et al., 2016b, Phys. Rev. Letters, 116, 241103

\noindent Ackermann, M., Ajello, M., Albert, A., et al., 2015, ApJL, 813, 41 (A15)

\noindent Aleksic J., Alvarez, E.A., Antonelli, L.A. et al., 2012, ApJ, 748, 46

\noindent Aleksic J., Ansoldi, S., Antonelli, L.A. et al., 2015, MNRAS, 450, 4399

\noindent Balokovic, M., Paneque, D., Madejski, G. et al., 2016, 819, 156

\noindent Baumgarte, T.W. \& Shapiro, S.L., 2011, Physics Today, 64, 32

\noindent Begelman, M.,  Blandford, R.D., \& Rees M.J., 1980, Nature, 207, 307

\noindent Boettcher, M. \& Dermer, C.D., 2002 ApJ, 564, 86

\noindent Burch, J.L., Torbert, R.B., Phan, T.D. et al., 2016, Science, 352, 1189

\noindent Cavaliere, A. \& D'Elia, V., 2002, ApJ, 571, 226

% \bibitem[]{} Coogan, Rosemary T.; Brown, Anthony M.; Chadwick, Paula M. , 2016,
% Localizing the gamma-ray emission region during the 2014 June
% outburst of 3C 454.3, MNRAS, 458, 354.

\noindent Celotti, A \& Ghisellini, G., 2008, MNRAS, 385, 283

\noindent Cohen, D.P, Romani, R.W., Filippenko, A.V., et al, 2014, ApJ, 797, 137

\noindent Colpi, M., 2014, Space Science Reviews, 183,  189

\noindent Coppi, B., 2016, Plasma Phys. Rep., 42, 383

\noindent Danforth, C.W., Keeney, B.A., Stocke, J.T., Shull, J.M., Yao, Y., 2010, ApJ, 720, 976

\noindent eLISA Consortium, 2013, The Gravitational Universe,  arXiv:1305.5720v1

\noindent Ghisellini, G., 2016, arXiv:1609.08606

%\bibitem[]{} {Giommi, P., 2015, JHEAp, 7, 173}

% \bibitem[]{} Hayashida, M., Nalewajko, K., Madejski, G.M. et al.,
% 2015, ApJ, 807, 79nel testo?)

\noindent Hovatta, T., Valtaoja, E., Tornikoski, M., Lahteenm\"aki, 2009,  A., 2009, A\&A 494, 527

% \bibitem[]{} Jenet, F., A. et al. 2004

\noindent Kagan, D., Sironi, L., Cerutti, B., Giannios, D., 2015, SSRv, 191, 545

\noindent Kelly. B. C., Bechtold, J. \& Siemiginowska, A., 2009, ApJ, 698, 895

%\bibitem[]{} {King, A ? ...}

\noindent Krawczynski, H., Hughes, S.B., Horan, D.,  et al., 2004, ApJ, 601, 151

%\bibitem[]{} MacDonald, N.R.,  Marscher, A. B., Jorstad, S. G., Joshi, M.,  2015,
%arXiv:1505.01239

\noindent Landau, L. \& Lifshitz, E.M., 1976,  \textit{Mechanics}, Oxford, Pergamon Press

\noindent Lapi, A., Raimundo, S., Aversa, R., et al., 2014,
ApJ, 782: 69

\noindent Maraschi, L., Ghisellini, G. \& Celotti, A., 1992, ApJ, 397, L5

\noindent Melzani, M., Walder, R., Folini, D., Winisdoerffer, C., Favre, J.M. et al., 2014, A\&A, 570A, 111

\noindent Mignone, A., Striani, E., Tavani, M., and Ferrari, A., 2013, MNRAS, 436, 1102

\noindent Nixon, C., \& King, A., 2015, arXiv:1505.07827

%\bibitem []{}Narayan, R., \& and Piran, T. 2012, MNRAS, 420, 604

%\bibitem []{} Paliya, V. et al. 2015 ...

\noindent Palenzuela, C., Garrett, T., Lehner, L., Liebling, S.L., 2010, PhRvD, 82, 404

\noindent Peterson, B.M., 1997, {\it An Introduction to Active Galactic Nuclei}, Cambridge Univ. Press

\noindent Peterson, B.M., 2006, {\it The Broad-Line Region in Active Galactic Nuclei}, Lect. Notes Phys. 693, 77 (Springer)

\noindent Raiteri, C. M., Stamerra, A., Villata, M., et al., 2015, MNRAS, 454, 353
% \bibitem []{} Raiteri, C. M. ... 2011

\noindent Rybicki, G.H., Lightman, A.P., 1979, {\it Radiative Processes in Astrophysics} (Wiley)

\noindent Sandrinelli, A., Covino, S., Treves, A.  2014, A \&A, 562, 79
%(... including a list of candidates ?)

%\bibitem []{} Sandrinelli, A., Covino, S., Dotti, M., Treves, A. 2016, AJ, 151, 54
%\bibitem []{} Sikora M., Rutkowski M. and Begelman M., 2015, MNRAS, submitted,
%arXiv:1511.08924v1

\noindent Sironi,L. \& Spitkovsky, A., 2014, ApJL, 783, 21

%\bibitem[]{}  Tavecchio, F. \& Ghisellini, G., 2008 ..

\noindent Striani, E., Mignone, A., Vaidya, B., Bodo, G., Ferrari, A., 2016, MNRAS, 462, 2970

\noindent Tavani, M., Vittorini, V., Cavaliere, A., 2015, ApJ, 814, 51

% \bibitem []{} Valtonen, M. J., Zola, S., Ciprini, S. et al., 2016 ApJL, 819, 37

\noindent Vaughan, S., Uttley, P., Markowitz, A.G., et al., 2016, MNRAS, 461, 3145

\noindent Volonteri, M., Bogdanovic, T., Dotti, M., Colpi, M.,
2015, Proc. XXIXth IAU General Assembly, arXiv:1509.09027v1

\noindent Yuan, Y, Nalewaiko, K., Zrake, J.,
East, W. E.,  Blandford, R., 2016, ApJ vol. 828; also preprint
arXiv:1604.03179.

%\end{thebibliography}   %{}

\newpage

% \section{Appendix - Deriving the Light Curves and the Corresponding Orbital Eccentricity}

 \appendix
 \section{Appendix:  Deriving the Light Curves and the Corresponding Orbital Eccentricities}

In Sect. 3  we  consider a gravitationally bound binary system
with  separation of several milli-pc.
 Drawing on time-honored dynamical relations
holding for Keplerian, elliptical orbits of the secondary (see,
e.g., Landau \& Lifshitz 1976, and their Figure 11) compared with
the light curves observed by A15 (cf, their Figs. 1, \, 2), {\mn
we detail how we deduce } our expected light curve and the
corresponding orbital \emph{eccentricity} $\varepsilon$.

We shall follow two ways. {\mn For semplicity, in both we neglect
the primary motion around the system's center of mass, and in
particular the size of its obit that  scales as $M_2/M_1 \lesssim
10^{-1}$}. We base on the gravitational stress from the secondary
being proportional to the  force between the two BHs, with
enhanced emission also proportional to it.

1) The ratio of the force at periastron to that at the opposite
vertex reads
\be F_{max}/F_{min} = (1 + \varepsilon)^2 / (1 -
\varepsilon)^2 . \label{eq-4} \en

In our approach, this also yields the ratio of gamma-ray fluxes at
maxima and minima.

From the gamma-ray light-curve in A15 (their Fig. 1)
%\ref{fig-app-1},
 we read out the observed value $\simeq 3$ for the
flux \emph{ratio }  between peaks and troughs. After Eq. 4
%\ref{eq-4}
 this corresponds to an orbital eccentricity
$\varepsilon \simeq 0.25$.

In addition, in Fig. 4 we plot the full dependence
 $f (t) \propto F [r(t)]$ of flux
 %
% \be F_{max}/F_{min} = (1 + \varepsilon)^2 / (1 -
%\varepsilon)^2 . \label{eq-5} \en
%
that  we expect along the orbit at a distance $r(t)$ from the
focus, parameterized with three
 eccentricity values: $\epsilon = 0.1,\, 0.2, \,  0.3$.
The data are overplotted, and agreement is found for $\epsilon =
0.2$

2) A second approach considers the time fraction along the orbit
\emph{ most} affected by such an effect. For an estimate,  we
concentrate on the points on the elliptical orbit marking the
intersections with the so-called \textit{latus rectum} $p$, i.e.,
the orbital distance corresponding to the "true anomaly" angle
$\phi = \pi/2$, with $\phi = 0$ at periastron. The time $\Delta
\,t$ spent on the orbital arc circling the focus from $p$ through
the periastron out to the opposite  phase,  is given by \be
\frac{\Delta \, t}{P} = \frac{\arccos (\varepsilon) - \varepsilon
\sqrt{ 1 - \varepsilon^2 }}{\pi} .\en
 Fig. 4
 %\ref{fig-app-2}
 shows the
ratio  $\Delta \,t / P$  as a function of orbital eccentricities.
From the observations recalled above
%in Fig. \ref{fig-app-1}
we see that the \emph{strongly}  enhanced emission lasts for $\sim
280$ days, corresponding to a fraction $\Delta \, t/P \simeq
0.35$. Interpreting this feature as caused by the gravitational
perturbation of a secondary BH in an elliptical orbit, we deduce
from Fig. 4
%\ref{fig-app-2}
 a value of the eccentricity $\varepsilon
\simeq 0.23$.

Thus we  conclude that on interpreting the repetitive pattern of
enhanced high-energy emissions from \pg in terms of stresses
induced by binary dynamics, both the above approaches lead to
evaluate for the orbital eccentricity  a value {\qq close to
$\varepsilon = 0.2$.}

Note that for a circular orbit the expected light curve is just
flat; in fact, from Eq. 4
%\ref{eq-4}
 one obtains  $F_{max}/F_{min} =1$. Concerning
Fig. 2 in the main text, we stress
% in Fig. \ref{fig-app-1}
that the dynamically \emph{stimulated} (as it were) emission
maxima largely exceed the \emph{spontaneous} random fluctuations
best noticed in the troughs.
%, even when such fluctuations are dynamically coordinated.

 \begin{figure}%
% \vspace*{-0.5cm}
%  \centerline{\includegraphics[width=15cm, angle = 0]{fig3-plus.jpg}}
%
 \centerline{\includegraphics[width=12cm,angle =-90]{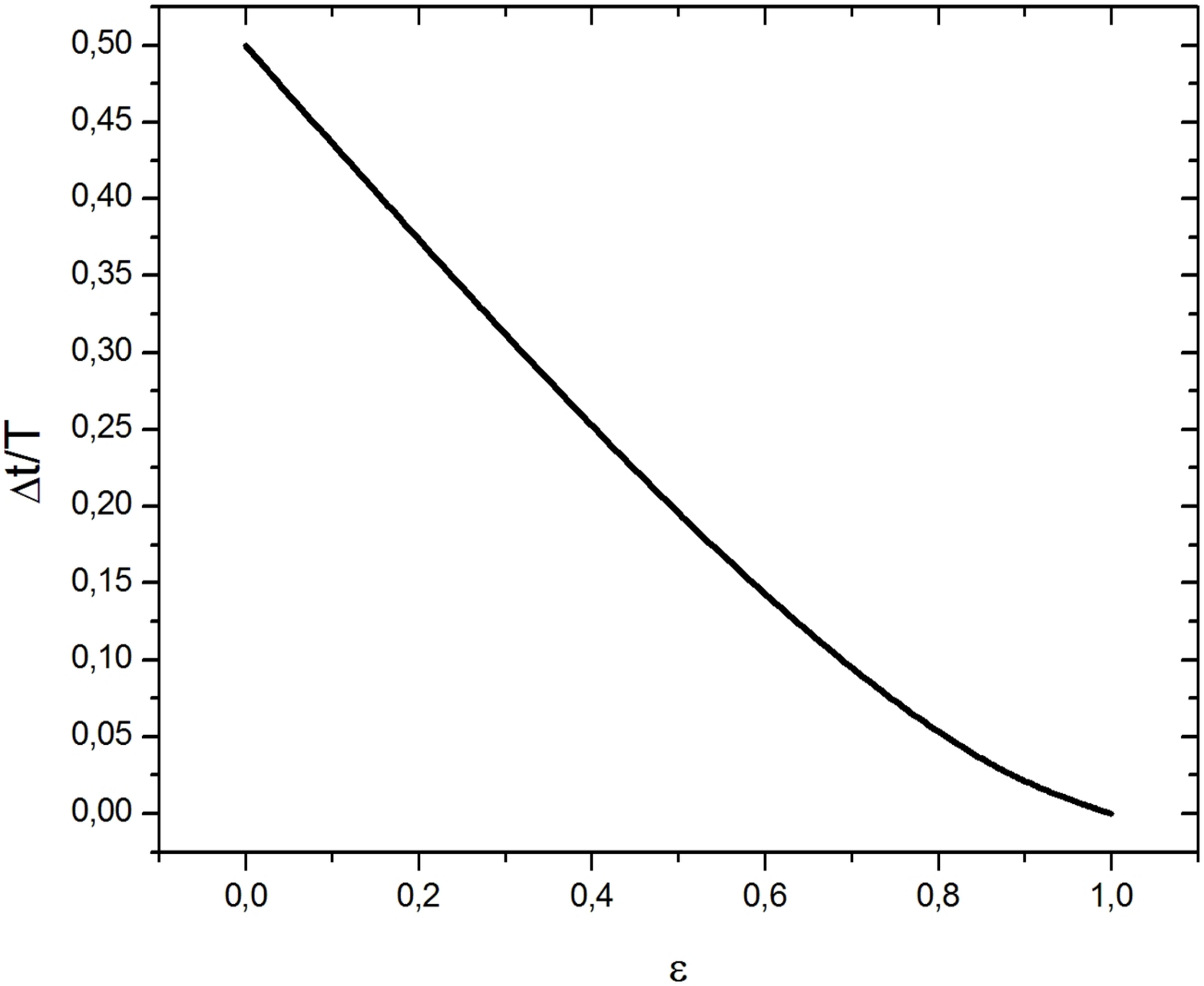} }  %{fig3nn.jpg}}
  \vspace*{-.2cm}
     \caption{ The ratio $\Delta \,t/P$  plotted as a function of the eccentricity
      for a Keplerian
     elliptical orbit. This quantity marks the strongly active fraction
     of the orbital
     period, that we take to be the interval between the intersections of
     the "latus rectum", each side of the  periastron.  }
 \label{fig-app-2}
 \end{figure}

\end{document}